\documentclass[%
 reprint,
superscriptaddress,
nolongbibliography,
 amsmath,amssymb,
 aps,
floatfix,
]{revtex4-2}

\usepackage{graphicx}
\usepackage{dcolumn}
\usepackage{bm}
\usepackage{hyperref}

\usepackage{siunitx} 
\usepackage{physics} 
\usepackage{cleveref} 

\usepackage{multirow} 
\usepackage{booktabs} 
\usepackage{mathtools}
\usepackage[caption=false]{subfig}

\usepackage{xcolor}
\hypersetup{
	colorlinks=true,
	citecolor=blue,
	linkcolor=blue,
	urlcolor=blue,
}

\Crefname{equation}{Eq.}{Eqs.}
\Crefname{figure}{Fig.}{Figs.}
\newcommand{\nuclide}[2]{\ensuremath{{}^{#1}\mathrm{#2}}}
\newcommand{\minv}[1]{\ensuremath{{({m}^{-1})}_{#1}}}

\begin{document}

\title{Neutron emission during fission and its impact on fission-fragment mass distribution studied by Langevin model}
\author{S.\,Takagi}
\affiliation{Graduate School of Science and Engineering, Kindai University, Higashi-Osaka, Osaka 577--8502, Japan}
\affiliation{Advanced Science Research Center, Japan Atomic Energy Agency, Tokai, Ibaraki 319--1195, Japan}
\author{S.\,Harada}
\author{Y.\,Aritomo}
\affiliation{Graduate School of Science and Engineering, Kindai University, Higashi-Osaka, Osaka 577--8502, Japan}
\author{K.\,Hirose}
\author{K.\,Nishio}
\affiliation{Advanced Science Research Center, Japan Atomic Energy Agency, Tokai, Ibaraki 319--1195, Japan}

\date{\today}
\begin{abstract}
  Actinide nuclei exhibit mass-asymmetric fission at low energy due to shell structure.
  The fission-fragment mass distributions produced at high energy tend to have a symmetric shape due to smearing of shell effects.
  On the other hand, the distribution can be changed by neutron emission occurring before fission, as this decreases the excitation energy of the fissioning nucleus, and thus revives the shell structure.
  In so called multichance fission, neutron emission is considered prior to fission at the initial nuclear shape, and competition between fission and neutron emission is determined with the framework of the statistical model.
  In the present work, we describe fission in the Langevin equations, and neutron emission is treated throughout the fission process.
  The calculation reproduces experimentally observed mass distributions, and for a wide range of initial compound-nucleus excitation energy up to 60 MeV.
  The results show that, while neutron emission dominates at the ground-state shape, it occurs along the shape evolution path to the scission point.
\end{abstract}

\maketitle

\section{Introduction}
Fission-fragment mass distributions (FFMDs) for low-energy fission of actinides from uranium to einsteinium show a double-humped shape, with the heavy fragment mass around $A_H = 140$\ \cite{Andreyev2018-ej,Britt1984-gf}.
These FFMDs have been interpreted to result from effects of nuclear shell structure\ \cite{Strutinsky1967-ua,Strutinsky1968-qf,Brack1972-qr,Wilkins1976-jb,Bjornholm1980-se,Brosa1990-ly}. 
Recently, theoretical calculation has reported that positions of the heavy fission fragment peak are regulated by proton shell structure of fragments around $Z = 52 \mbox{--} 56$, which prefer octupole deformation\ \cite{Scamps2018-xe, Schmidt2001-fz, Mahata2022-nk}. 

For a highly excited compound nucleus, the mass-asymmetric shape of the FFMD tends to disappear due to the quenching of shell structure\ \cite{Gonnenwein1991-mv}.
However, the measured FFMD from a highly excited nucleus is influenced by the effects of multi-chance fission, i.e. fission after the evaporation of neutrons\ \cite{Ryzhov2011-mh,Naik2012-wk,Nadtochy2014-xh,Khuyagbaatar2015-ou,Duke2016-cr,Moller2017-ss,Hirose2017-ns, Tanaka2019-yy, Vermeulen2020-lf,Lovell2021-uu,Berriman2022-ng,Santra2023-oh,Pomorski2024-hr,Liu2025-zy,Ivanyuk2025-oi}.
In such a case, the shell structure is revived through the removal of excitation energy by neutrons, which restores an asymmetric distribution.

To determine the probability of multichance fission, a statistical model framework has been adopted\ \cite{Schmidt2016-ps,Vandenbosch1973-ev}.
In this model neutron evaporation can happen only when a nucleus is in its ground-state shape, i.e. before the excess energy drives the nucleus to deform toward scission.
Such neutron emission affects the overall potential energy surface due to changes in the shell correction energies.
In this study, we implemented a dynamical model (based on the Langevin equations) in which neutrons can be evaporated at any stage of the fission process, and in which neutron emission competes with nuclear shape change as a means of dissipating excess energy.
In our model, potential energy changes, resulting from updated shell corrections, are also made at any point where neutron evaporation occurs.
We found that this new procedure can also reproduce the measured FFMDs for high-energy fission. 
In contrast to the conventional method, this approach has the following advantages.
First of all, we do not have to separately calculate neutron emission and fission with different models.
This reduces the cost of computation encountered in the traditional approach, in which several calculations must be performed in advance when changing the initial excitation energy.
Secondly, the new calculation does not require nuclide-dependent parameters, such as fission barrier height and level density parameters, which are mandatory in the traditional statistical model.
This reduces the burden on preparing reliable inputs and creates a versatility to calculate fission of unexplored region on the chart of nuclides.
Finally, our new approach of neutron evaporation during fission can be used to discuss the time for the fission process by comparing the calculation with the experimental data of the number of pre-scission neutrons\ \cite{Hinde1992-ss, Hinde2021-md}.

\begin{figure*}[htbp]
  \hspace{-3cm}
  \subfloat{
    \includegraphics[width=0.7\linewidth]{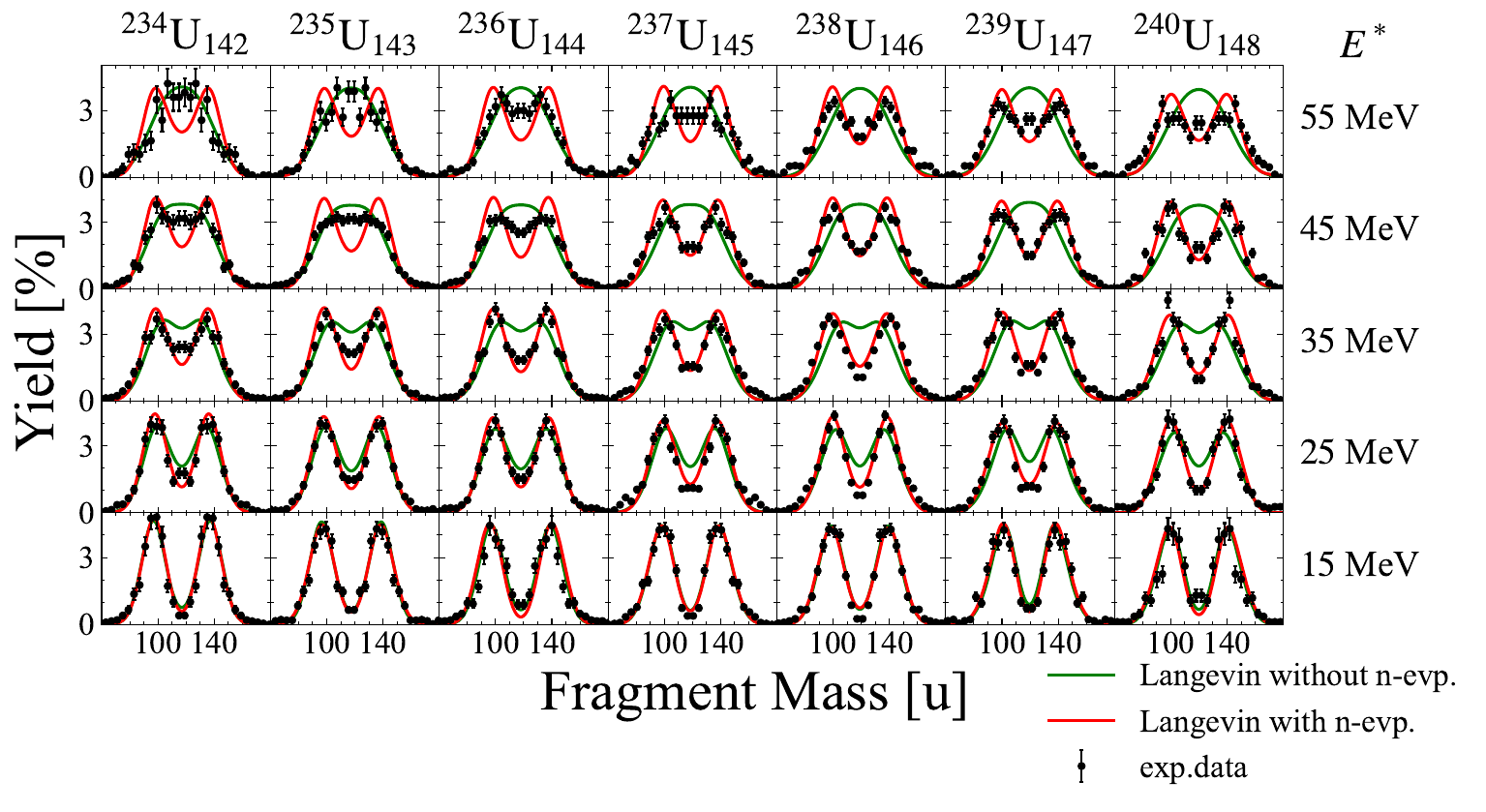}\label{fig:U234_240}
  }
  \\

  \subfloat{
    \includegraphics[width=0.7\linewidth]{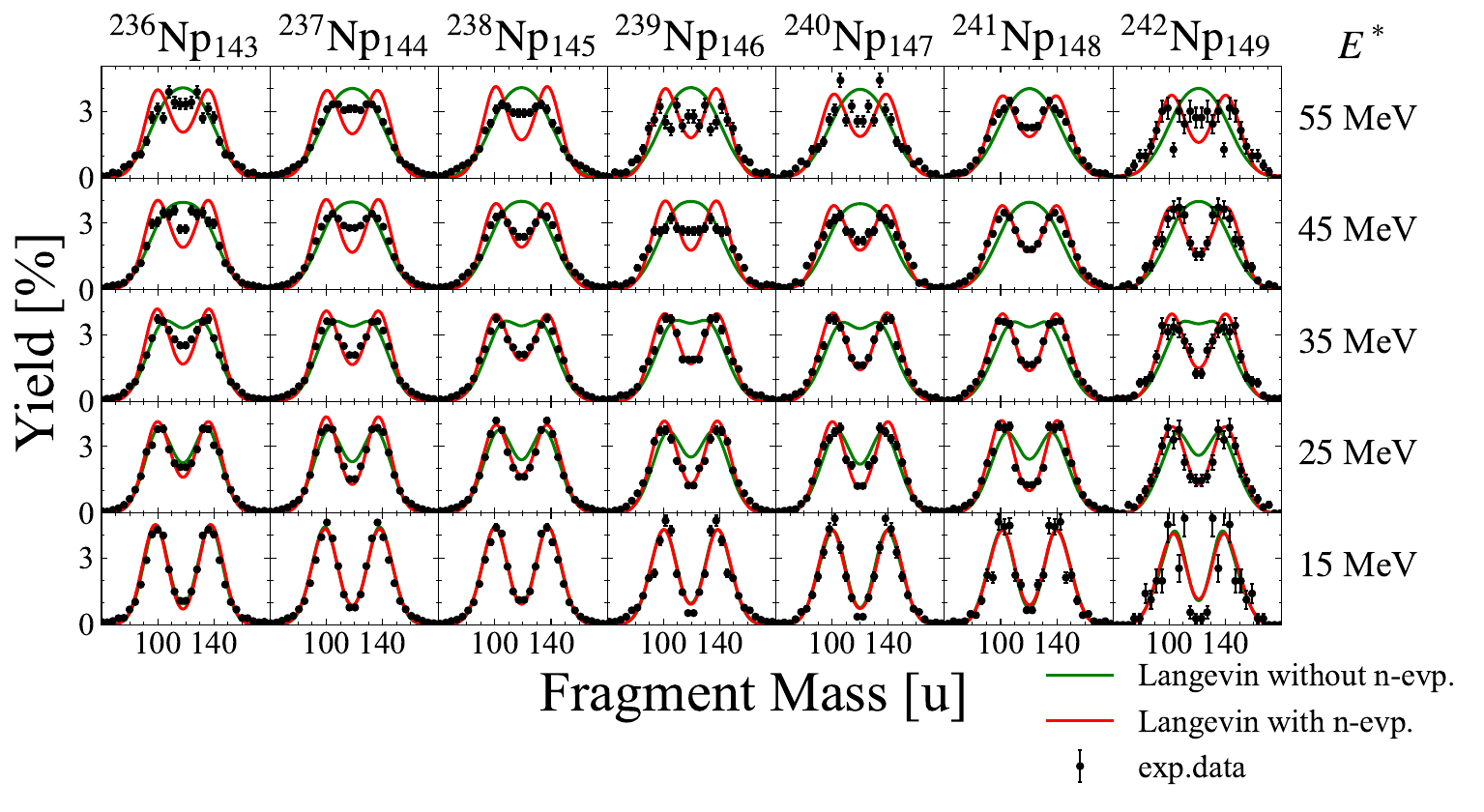}\label{fig:Np236_242}
  }
  \\
  \hspace{3cm}
  \subfloat{
    \includegraphics[width=0.7\linewidth]{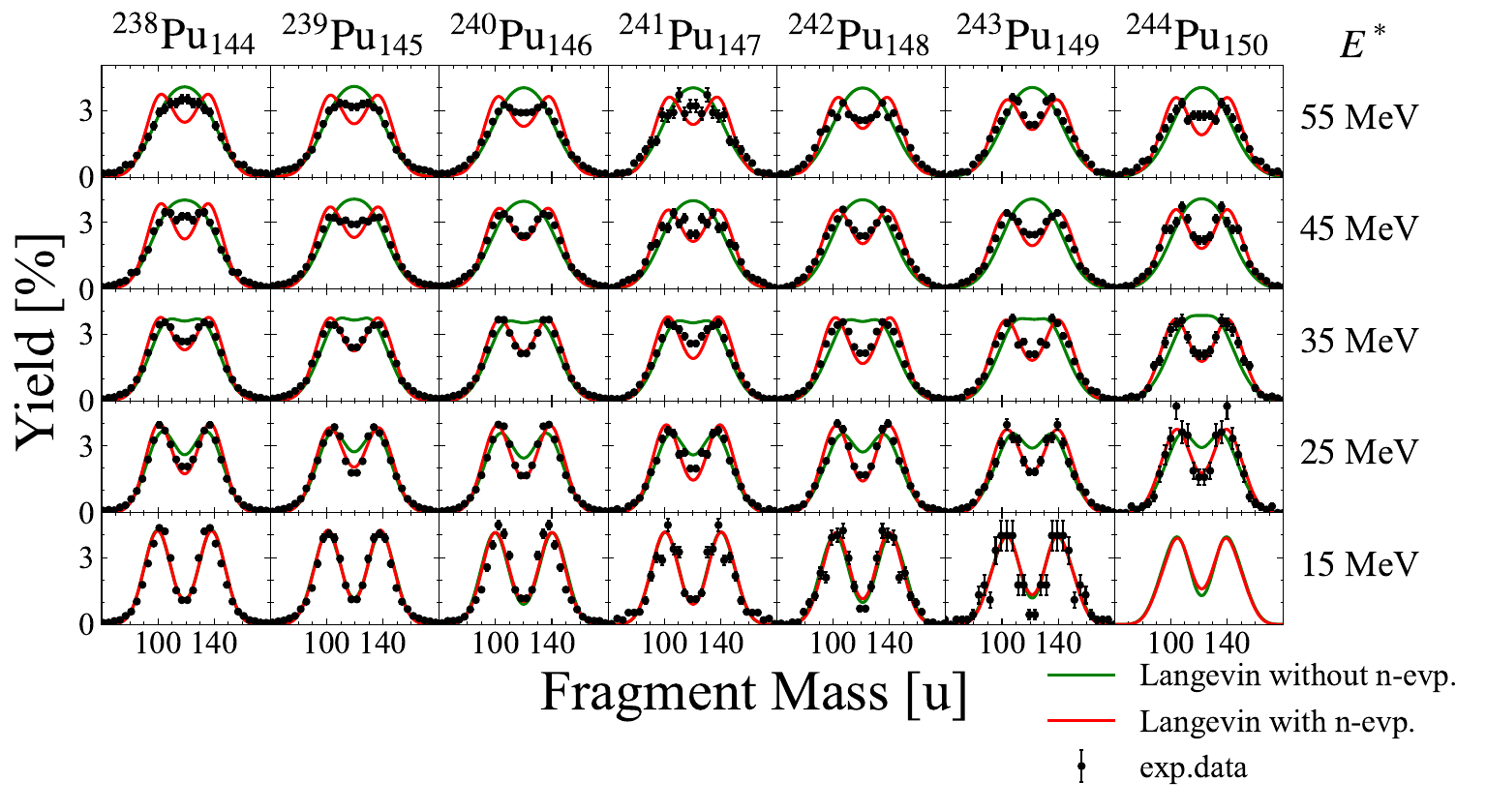}\label{fig:Pu238_244}
  }
  \caption{
        Comparison of the measured FFMDs\ \cite{Hirose2017-ns, Vermeulen2020-lf} (black dots with error bars) for uranium, neptunium, and plutonium isotopes with the Langevin calculation.
        The red and green solid lines are the results with and without neutron emission during fission, respectively.
        The figure is arranged so that the isotones of the three compound nuclei species are vertically aligned.
        Average excitation energies are shown on the right side of the figure. 
    }
  \label{fig:FFMD}
\end{figure*}

\section{Model}
\subsection{Dynamical model}
We used three-dimensional Langevin equations to calculate the time evolution of the shape of the nucleus.
Although the present calculation has fewer dimensions than comparable 4D \cite{Nadtochy2014-xh,Pahlavani2016-ga,Ishizuka2017-ec,Sierk2017-ih,Usang2019-gb,Pomorski2024-hr} and 5D \cite{Ivanyuk2025-oi} Langevin model used in the description of fission,
our calculation successfully reproduces the fission-fragment mass distribution of the lowest excitation energies, as will be shown in \Cref{fig:FFMD}.
In our model, the nuclear shape is defined by the two-center parametrization\ \cite{Maruhn1972-rv, Sato1978-jy}, which has three deformation parameters, $z_0$, $\delta$, and $\alpha$.
Here, $z_0$ is the distance between the center of two potentials, $\delta$ denotes the deformation parameter of each fragment ($\delta = \delta_1 = \delta_2$), and $\alpha$ is the mass asymmetry of the fragments.
The parameter $\delta$ is defined as $\delta = 3(a - b)/(2a + b)$ using the half-length of the ellipse axes $a$ (symmetry-axis direction) and $b$ (radial direction).
The mass asymmetry $\alpha = (A_1 - A_2)/(A_1 + A_2)$ is given by the fragment mass $A_1$ and $A_2$.
In addition, to reduce computation time we introduce the scaled coordinate $z$, defined as $z = z_0/(R_{\mathrm{CN}}B)$,
where $R_{\mathrm{CN}}$ is the radius of the spherical compound nucleus and $B = (3 + \delta)/(3 - 2\delta)$.
These three collective coordinates are abbreviated as $q$, with $q = \{z, \delta, \alpha \}$.
The neck parameter $\varepsilon$ is defined as
\begin{equation}
  \varepsilon(A_{\mathrm{CN}} ) = 0.01007A_{\mathrm{CN}} - 1.94,
\end{equation}
where $A_{\mathrm{CN}}$ is the mass number of the compound nucleus \cite{Miyamoto2019-gd}.

In the fission process, the nuclear potential is described by the adiabatic potential $V (q)$.
The potential energy $V$ is calculated by
\begin{equation} \label{eq:potential}
  \begin{aligned}
    V(q,T)                 & = V_{\mathrm{LDM}}(q) + V_{\mathrm{SH}}(q,T),                 \\
    V_{\mathrm{LDM}}(q)    & = E_\mathrm{S}(q)+E_\mathrm{C}(q),                            \\
    V_{\mathrm{SH}}(q,T)   & = E_{\mathrm{SH}}^{0}(q)\Phi(T),                              \\
    E_{\mathrm{SH}}^{0}(q) & = \Delta{E_{\mathrm{shell}}}(q) + \Delta{E_\mathrm{pair}}(q), \\
    \Phi(T)                & = \exp(-\frac{a_n T^2}{E_d}).
  \end{aligned}
\end{equation}
In \Cref{eq:potential}, $V_{\mathrm{LDM}}$ is the potential energy from the finite-range liquid drop model\ \cite{Krappe1979-nk},
given as a sum of the surface energy $E_\mathrm{S}(q)$ and the Coulomb energy $E_\mathrm{C}(q)$.
The microscopic potential, $V_{\mathrm{SH}}$, is calculated at $T=0$ as the sum of the shell correction energy $\Delta{E_{\mathrm{shell}}}(q)$, evaluated by the Strutinsky method\ \cite{Strutinsky1967-ua, Strutinsky1968-qf},
and the pairing correction energy $\Delta{E_\mathrm{pair}}(q)$, estimated by the Bardeen-Cooper-Schrieffer (BCS) method\ \cite{Nilsson1969-gc, Brack1972-qr}.
$T$ is the temperature of the nucleus calculated from the intrinsic excitation energy of the composite system, represented by \Cref{eq:intrinsic} below.
The temperature dependence factor $\Phi(T)$, which includes the level-density parameter $a_n$ [see \Cref{eq:level_densty_parameter} below], is discussed in detail in Ref.\ \cite{Aritomo2004-tj}.
The shell-damping energy $E_\mathrm{d}$ was chosen as $\SI{20}{\MeV}$, as proposed by Ignatyuk \textit{et al.}\ \cite{Ignatyuk1975-iw}.

The Langevin equation\ \cite{Aritomo2004-tj} is given as
\begin{equation} \label{eq:langevin}
  \begin{aligned}
    \dv{q_i}{t} & = {\qty(m^{-1})}_{ij}p_{j},                                    \\
    \dv{p_i}{t} & = -\pdv{V}{q_{i}} - \frac{1}{2}\pdv{q_{i}} \minv{jk}p_{j}p_{k} \\
                & \phantom{{}={}} - \gamma_{ij}\minv{jk}p_{k} + g_{ij}R_{j}(t),
  \end{aligned}
\end{equation}
where $q_i = \{z, \delta, \alpha\}$, and $p_i = m_{ij}\mathrm{d}{q_j}/\mathrm{d}{t}$ ($i$ represents $z, \delta, \alpha$) denotes the momentum conjugated to $q_i$.
In \Cref{eq:langevin}, $m_{ij}$ and $\gamma_{ij}$ are the shape-dependent collective inertia and one-body friction tensors, respectively.
We adopted the hydrodynamical inertia tensor $m_{ij}$ from the Werner-Wheeler approximation\ \cite{Davies1976-yi}.
The tensor $\gamma_{ij}$ is calculated with the wall-and-window formula\ \cite{Randrup1984-pc, Sierk1980-im}.
The normalized random force tensor $R_{j}(t)$ is assumed to be white noise, $\ev{R_{j}(t)} = 0$ and $\ev{R_{i}(t_{1}) R_{j}(t_{2})} = 2\delta_{ij}\delta(t_{1} - t_{2})$.
The strength of the random force, $g_{ij}$, is related to the friction tensor by the classical Einstein relation,
\begin{equation} \label{eq:einstein_relation}
  \sum_{k}g_{ik}g_{jk} = \gamma_{ij}T.
\end{equation}
The temperature $T$ is related to the intrinsic excitation energy, $E_{\mathrm{int}}$, of the composite system.
$E_{\mathrm{int}}$ is calculated at each step of the trajectory calculation by,
\begin{equation}
  \begin{aligned}
    E_{\mathrm{int}} & = E^* - \frac{1}{2}\minv{ij}p_{i}p_{j} - V(q, T=0), \\
                     & = a_nT^2, \label{eq:intrinsic}
  \end{aligned}
\end{equation}
where $E^*$ is the excitation energy at each step of the nuclear shape and/or neutron emission.

\subsection{Neutron-decay width}
Neutron emission during fission was introduced into the Langevin calculation following the prescription of Ref.\ \cite{Wada1993-zg, Frobrich1993-mj}.
Following each neutron emission, potential and excitation energy are changed because of the change of the fissioning isotope and removal of excitation energy.
When a neutron is emitted during the time evolution of the nuclear shape, the potential energy changes due to several effects.
One such effect comes from an increase in the shell correction energy, due to a decrease in the excitation energy following neutron emission (as discussed in Ref. \cite{Tanaka2019-yy}).
Another arises from the change in the number of neutrons in the nucleus, as the neutron emission alters the nuclide to ($Z,N$-1).
Thus, the potential energy is calculated for the fissioning nucleus ($Z,N$-1).

For each fixed time step, $\Delta t$, of the calculation, the change of the nuclear-shape coordinate $q$ is determined.
Also for each step the probability of neutron emission is calculated using the lifetime of neutron emission $\tau_n$, 
\begin{equation}
  \frac{\Delta t}{\tau_{n}} \geq \zeta, \label{eq:competition_n_langevin}
\end{equation}
where the time step and the lifetime are defined as $\Delta t \leq \tau_n$\ \cite{Frobrich1993-mj}.
Here, $\tau_n$ is given by $\tau_n = \hbar/\varGamma_{n}$ using the neutron-decay width $\varGamma_n$.
The parameter $\zeta$ is a uniform random number satisfying $0\leq \zeta\leq 1$.

A simple description of the neutron-decay width based on detailed balance is adopted \cite{Weisskopf1937-tb,Bohr1939-jv}.
\begin{equation}
    \begin{aligned}
        \varGamma_n^J  = &\frac{1}{2\pi \rho(U^*,J)}g_n \int_{0}^{U^*-B_n} \\
        &\times \sum_{\ell} (2\ell+1) T_l(E_n)\rho(U^* - B_n - E_n, \ell) \dd{E_n},
    \end{aligned}
\end{equation}
where $U^*$ is the effective excitation energy of the parent nucleus, defined as $U^* = E_{\mathrm{int}} - E_\delta$, in which $E_\delta$ indicates the pairing energy represented as
\begin{equation}
    E_\delta = 
    \begin{cases}
        24/\sqrt{A}  & (Z\ \&\ N\ \text{even}), \\
        12/\sqrt{A}  & (A\ \text{odd}), \\
        0  &           (Z\ \&\ N\ \text{odd}).
    \end{cases}
\end{equation}
$T_l$ is the transmission coefficient of the neutron for the centrifugal potential.
$g_n$ is defined as $g_n = 2s_n + 1 = 2$ ($s_n$ is the spin of the neutron).
The symbol $E_n$ is the kinetic energy of the evaporated neutron.
The effective excitation energy of the nucleus can be expressed by $U^* - B_n - E_n$,
where $B_n$ is the neutron separation energy.

The level density $\rho$ is calculated using the Fermi-gas model \cite{Mughabghab1998-gw}.
The spin $J$-dependent level density $\rho(U^*, J)$ is given as

\begin{equation}
  \begin{aligned}
    \rho(U^*,J) & = \frac{\sqrt{\pi}}{12 a_n^{1/4}U^{*5/4}}\exp(2\sqrt{a_nU^*})                                     \\
                & \phantom{{}={}} \times \frac{2J + 1}{2\sqrt{2\pi}\sigma^3}\exp{-\frac{(J+1/2)^2}{2\sigma^2}}.
  \end{aligned}
\end{equation}
Here, $\sigma$ is the spin cut-off parameter defined as $\sigma^2 = IT/\hbar^2$, where $I$ and $T$ are the moment of inertia and temperature, respectively.
The level-density parameter $a_n$ is given as
\begin{equation}
  a_n = \tilde{a}_n \left[1 + \frac{E^0_\mathrm{SH}}{U^*} \left(1 - \exp(- \frac{U^*}{E_d} )\right) \right]. \label{eq:level_densty_parameter}
\end{equation}
The coefficient $\tilde{a}_n$ of the level-density parameter $a_n$ takes into account the deformation of the nucleus\ \cite{Toke1981-pn},
\begin{equation}
  \begin{aligned}
    \tilde{a}_n & = \frac{A}{14.61} (1 + 3.114A^{-1/3} X_{22} + 5.626A^{-2/3} X_{33} ),     \\
    X_{22}      & =
    \begin{dcases}
     2\pi \qty(a b \frac{\sin^{-1}\phi}{\phi} b^2 ) / 4 \pi R_{\mathrm{CN}}^2  & (\delta > 0),\\
     1  & (\delta \le 0),
    \end{dcases} \\
    X_{33}      & =
    \begin{dcases}
        \frac{2\pi}{a\phi} \qty(\frac{b^2}{2} \log\frac{1 + \phi}{1 - \phi} ) /8\pi R_{\mathrm{CN}}  & (\delta > 0), \\
        0.5  & (\delta \le 0),
    \end{dcases} \\ 
    \phi        & = \sqrt{1 - \frac{1}{\eta^2}}, \eta = a / b.
  \end{aligned}
\end{equation}
Concerning the value of $a_n$ through the fission process, by definition it does not change for oblate shape of $\delta < 0$, and changes only slightly for prolate deformation, varying by less than 4\% up to the largest elongation $\delta = 0.8$, 
which is the practical maximum deformation reached in our Langevin calculation, see \Cref{fig:neutron_map} (a). 
We note that we do not introduce the enhancement of level density due to collective motion \cite{Gavron1976-ra,Dossing2019-vw}.

\begin{figure*}[htbp]
  \centering
  \subfloat{
    \includegraphics[width=0.9\linewidth]{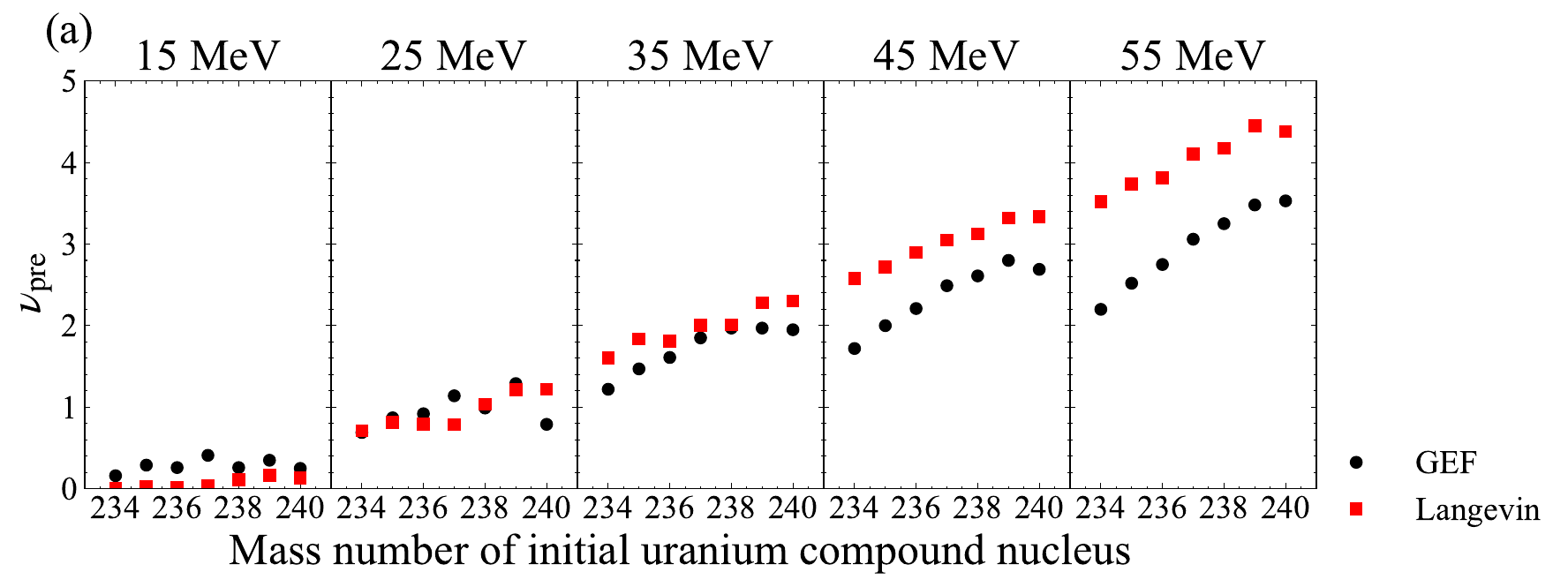}\label{fig:average_neutron_92}
  }
  \\
  \subfloat{
    \includegraphics[width=0.9\linewidth]{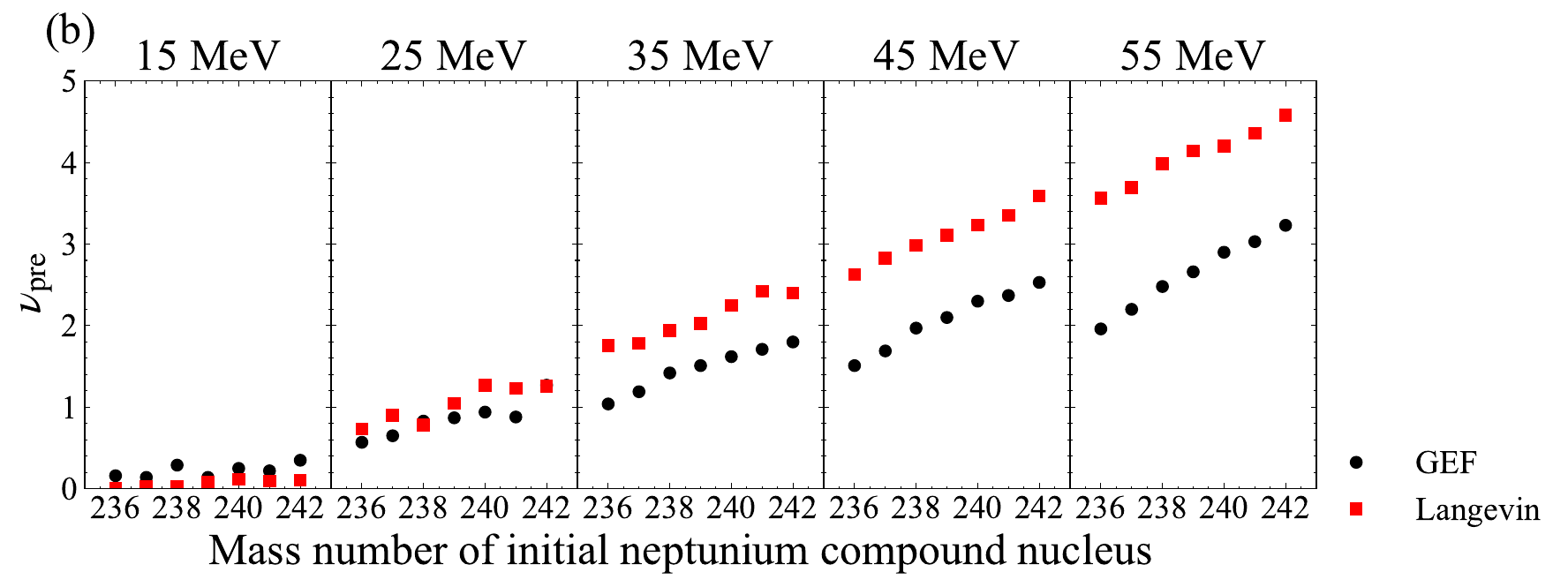}\label{fig:average_neutron_93}
  }
  \\
  \subfloat{
    \includegraphics[width=0.9\linewidth]{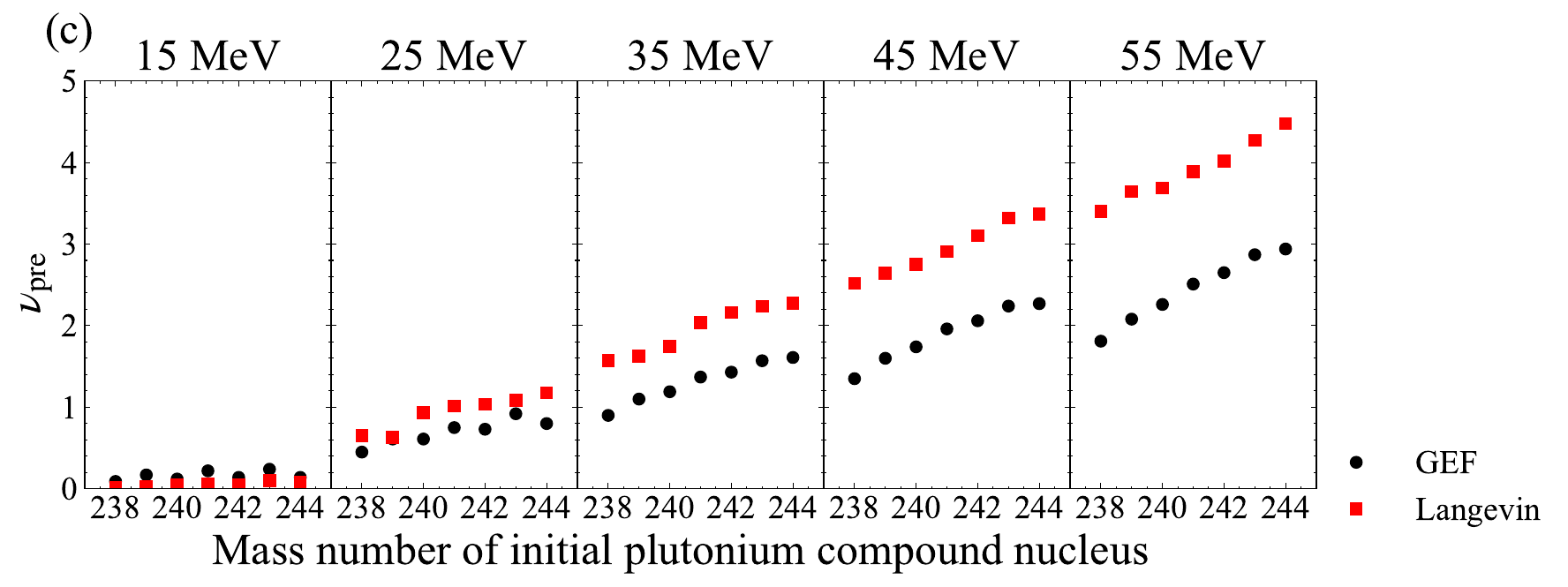}\label{fig:average_neutron_94}
  }
  \caption{
    The number of neutron emission $\nu_\mathrm{pre}$ of (a) $\nuclide{234-240}{U}$, (b) $\nuclide{236-242}{Np}$, and (c) $\nuclide{238-244}{Pu}$ with the initial excitation energy $E^* = \qtyrange[range-units=single]{15}{55}{\MeV}$ calculated by the present Langevin model (red square).
    Values from GEF code (black circle) are also shown.
  }\label{fig:average_neutron}
\end{figure*}

\section{Result and Discussion}
Figure \ref{fig:FFMD} shows the FFMDs for 21 nuclides ($\nuclide{234-240}{U}$, $\nuclide{236-242}{Np}$, and $\nuclide{238-244}{Pu}$) with an average initial excitation energy range from $E^*= 15$ to \SI{55}{\MeV} (\SI{10}{\MeV} bin).
The solid lines show the calculated mass distribution without neutron emission (green) and with neutron emission (red), respectively.
The data points with error bars are the experimental FFMDs measured in multinucleon-transfer reactions of $\nuclide{18}{O} + \nuclide{238}{U}$ and $\nuclide{18}{O}+\nuclide{237}{Np}$, taken at the JAEA tandem accelerator facility\ \cite{Hirose2017-ns,Vermeulen2020-lf}. 
At $E^* = \SI{15}{MeV}$, sufficient data for $\nuclide{244}{Pu}$ were not available to obtain the FFMD, so only calculated values are shown.
At low excitation energies of \SI{15}{MeV}, for all nuclides, the experimental data showing double-humped mass asymmetric fission are well reproduced by both calculations.
The effects of neutron-emission in fission is more significant at higher excitation energies.
At excitation energy larger than $E^*=\SI{40}{\MeV}$, calculations without neutron emission show a single Gaussian shape due to the disappearance of shell effects.
On the contrary, the calculation which introduces neutron emission maintains the double-humped structure, even at the highest excitation energy ($E^*=\SI{55}{\MeV}$), and the experimental data are better described by this model.
The calculation also captures the change of the FFMD with proton number of the compound nucleus at high excitation energy.
For example, at neutron number $N=148$, the ratio of the peak yield of the mass-asymmetric fission to that of the symmetric valley (peak-to-valley ratio) decreases from uranium to plutonium.
However, the agreement of the present calculation with the experimental data is not as good as the previous calculation in Ref.\cite{Tanaka2019-yy} based on the GEF code \cite{Schmidt2016-ps}.
At high energy, from \SI{45}{\MeV} to \SI{55}{\MeV}, the present calculation shows larger peak-to-valley ratio than the experimental data, especially for neutron-deficient fissioning nuclei.
This is because the present calculation generates a larger number of neutrons than the GEF code, thus revival of shell effects is more significant (see discussion below).
Conversely, the calculation in Ref.\cite{Tanaka2019-yy} agrees well with the experimental data, including the isotope dependence of the FFMD at high energy, due to developing larger peak-to-valley ratio toward neutron-rich isotopes.

Figure \ref{fig:average_neutron} shows the calculated number of neutrons emitted before the scission point ($\nu_{\mathrm{pre}}$) for (a) U, (b) Np, and (c) Pu isotopes, with excitation energies from 15 to \SI{55}{MeV}.
Our results are shown by squares (red).
We also show the number of neutrons emitted while in the ground-state shape, calculated by GEF (2016/1.2) (black circle), as used in\ \cite{Hirose2017-ns, Tanaka2019-yy, Vermeulen2020-lf}.
For all the studied nuclei, the number of neutrons emitted increases linearly with the excitation energy of the initial compound nucleus.
We note that the average kinetic energy of emitted neutrons in the present model is \SI{1.64}{\MeV}, which is very close to the value of \SI{1.9}{\MeV} obtained in the previous calculations \cite{Tanaka2019-yy}.
In \Cref{fig:average_neutron}, there are data points for which the neutron multiplicity drops slightly with increasing fissioning nucleus mass number.
This comes from fluctuation of when and how neutrons are emitted during fission.
Our results of the total number of neutrons emitted is almost the same as the GEF calculation up to $E^* = \SI{35}{\MeV}$ for uranium and up to $\SI{25}{\MeV}$ for neptunium/plutonium nuclides.
Beyond these energies, the present calculation generates more neutrons.
In addition, focusing on the number of neutrons at each excitation energy, the number of neutrons increases as the mass number of the compound nucleus increases.
This is because of the lowering of neutron binding energy toward heavier isotopes (neutron-rich isotopes), thus a larger probability of neutron emission $(\nu_\mathrm{pre})$.
In \Cref{table:cal_result}, we summarize the number of neutrons emitted in the present calculation and two neutron separation energies $(S_{2n})$ for initial compound nuclei\ \cite{Moller2016-cy}.

\setlength{\tabcolsep}{8pt}
\renewcommand{\arraystretch}{1.0}
\begin{table*}[hbtp]
  \caption{
    The calculated number of neutrons emitted $\nu_{\mathrm{pre}}$ for initial compound nucleus $\nuclide{234-240}{U}$, $\nuclide{236-242}{Np}$, and $\nuclide{238-244}{Pu}$ with initial-excitation energy $E^*$ of \qtyrange[range-phrase=~--~, range-units=single]{15}{55}{\MeV}.
    $S_{2n}$ is the two neutron separation energy for initial compound nuclei\ \cite{Moller2016-cy}.
  }\label{table:cal_result}
  \centering
  \begin{tabular}{lccccccccccc}
    \toprule
                                      &$E^*$ & \multicolumn{1}{c}{$S_{2n}$} &  \multicolumn{1}{c}{$\nu_\mathrm{pre}$} &                                    & $E^*$ & \multicolumn{1}{c}{$S_{2n}$} &  \multicolumn{1}{c}{$\nu_\mathrm{pre}$} &                                    & $E^*$ & \multicolumn{1}{c}{$S_{2n}$} & \multicolumn{1}{c}{$\nu_\mathrm{pre}$} \\
                                      & [MeV] & [MeV] &                              &                                    & [MeV] & [MeV] &                              &                                    & [MeV] & [MeV] &                             \\
    \midrule
    \multirow{5}{*}{\nuclide{234}{U}} & 15    &\multirow{5}{*}{12.60} &  0.007                       & \multirow{5}{*}{\nuclide{236}{Np}} & 15    &\multirow{5}{*}{12.72} &  0.010                       & \multirow{5}{*}{\nuclide{238}{Pu}} & 15    &\multirow{5}{*}{12.88} & 0.011                       \\
                                      & 25    & &  0.712                       &                                    & 25    & &  0.733                       &                                    & 25    & & 0.650                       \\
                                      & 35    & &  1.601                       &                                    & 35    & &  1.755                       &                                    & 35    & & 1.572                       \\
                                      & 45    & &  2.576                       &                                    & 45    & &  2.627                       &                                    & 45    & & 2.519                       \\
                                      & 55    & &  3.519                       &                                    & 55    & &  3.562                       &                                    & 55    & & 3.401                       \\
    \midrule
    \multirow{5}{*}{\nuclide{235}{U}} & 15    &\multirow{5}{*}{12.14} &  0.021                       & \multirow{5}{*}{\nuclide{237}{Np}} & 15    &\multirow{5}{*}{12.31} &  0.023                       & \multirow{5}{*}{\nuclide{239}{Pu}} & 15    &\multirow{5}{*}{12.64} & 0.028                       \\
                                      & 25    & &  0.815                       &                                    & 25    & &  0.903                       &                                    & 25    & & 0.633                       \\
                                      & 35    & &  1.838                       &                                    & 35    & &  1.784                       &                                    & 35    & & 1.626                       \\
                                      & 45    & &  2.720                       &                                    & 45    & &  2.824                       &                                    & 45    & & 2.646                       \\
                                      & 55    & &  3.734                       &                                    & 55    & &  3.693                       &                                    & 55    & & 3.646                       \\
    \midrule
    \multirow{5}{*}{\nuclide{236}{U}} & 15    &\multirow{5}{*}{11.84} &  0.013                       & \multirow{5}{*}{\nuclide{238}{Np}} & 15    &\multirow{5}{*}{12.06} &  0.031                       & \multirow{5}{*}{\nuclide{240}{Pu}} & 15    &\multirow{5}{*}{12.18} & 0.049                       \\
                                      & 25    & &  0.791                       &                                    & 25    & &  0.781                       &                                    & 25    & & 0.935                       \\
                                      & 35    & &  1.810                       &                                    & 35    & &  1.941                       &                                    & 35    & & 1.742                       \\
                                      & 45    & &  2.899                       &                                    & 45    & &  2.983                       &                                    & 45    & & 2.754                       \\
                                      & 55    & &  3.814                       &                                    & 55    & &  3.987                       &                                    & 55    & & 3.688                       \\
    \midrule
    \multirow{5}{*}{\nuclide{237}{U}} & 15    &\multirow{5}{*}{11.67} &  0.041                       & \multirow{5}{*}{\nuclide{239}{Np}} & 15    &\multirow{5}{*}{11.70} &  0.078                       & \multirow{5}{*}{\nuclide{241}{Pu}} & 15    &\multirow{5}{*}{11.77} & 0.059                       \\
                                      & 25    & &  0.786                       &                                    & 25    & &  1.049                       &                                    & 25    & & 1.012                       \\
                                      & 35    & &  2.004                       &                                    & 35    & &  2.024                       &                                    & 35    & & 2.035                       \\
                                      & 45    & &  3.047                       &                                    & 45    & &  3.108                       &                                    & 45    & & 2.907                       \\
                                      & 55    & &  4.102                       &                                    & 55    & &  4.143                       &                                    & 55    & & 3.888                       \\
    \midrule
    \multirow{5}{*}{\nuclide{238}{U}} & 15    &\multirow{5}{*}{11.28} &  0.113                       & \multirow{5}{*}{\nuclide{240}{Np}} & 15    &\multirow{5}{*}{11.28} &  0.115                       & \multirow{5}{*}{\nuclide{242}{Pu}} & 15    &\multirow{5}{*}{11.55} & 0.046                       \\
                                      & 25    & &  1.035                       &                                    & 25    & &  1.266                       &                                    & 25    & & 1.034                       \\
                                      & 35    & &  2.012                       &                                    & 35    & &  2.247                       &                                    & 35    & & 2.161                       \\
                                      & 45    & &  3.124                       &                                    & 45    & &  3.233                       &                                    & 45    & & 3.104                       \\
                                      & 55    & &  4.176                       &                                    & 55    & &  4.203                       &                                    & 55    & & 4.016                       \\
    \midrule
    \multirow{5}{*}{\nuclide{239}{U}} & 15    &\multirow{5}{*}{10.96} &  0.165                       & \multirow{5}{*}{\nuclide{241}{Np}} & 15    &\multirow{5}{*}{11.14} &  0.095                       & \multirow{5}{*}{\nuclide{243}{Pu}} & 15    &\multirow{5}{*}{11.34} & 0.103                       \\
                                      & 25    & &  1.217                       &                                    & 25    & &  1.231                       &                                    & 25    & & 1.087                       \\
                                      & 35    & &  2.279                       &                                    & 35    & &  2.421                       &                                    & 35    & & 2.236                       \\
                                      & 45    & &  3.322                       &                                    & 45    & &  3.351                       &                                    & 45    & & 3.319                       \\
                                      & 55    & &  4.449                       &                                    & 55    & &  4.358                       &                                    & 55    & & 4.270                       \\
    \midrule
    \multirow{5}{*}{\nuclide{240}{U}} & 15    &\multirow{5}{*}{10.73} &  0.132                       & \multirow{5}{*}{\nuclide{242}{Np}} & 15    &\multirow{5}{*}{11.04} &  0.104                       & \multirow{5}{*}{\nuclide{244}{Pu}} & 15    &\multirow{5}{*}{11.05} & 0.080                       \\
                                      & 25    & &  1.218                       &                                    & 25    & &  1.259                       &                                    & 25    & & 1.178                       \\
                                      & 35    & &  2.304                       &                                    & 35    & &  2.400                       &                                    & 35    & & 2.277                       \\
                                      & 45    & &  3.334                       &                                    & 45    & &  3.590                       &                                    & 45    & & 3.369                       \\
                                      & 55    & &  4.382                       &                                    & 55    & &  4.578                       &                                    & 55    & & 4.476                       \\
    \bottomrule
  \end{tabular}
\end{table*}

\begin{figure}
  \centering
  \subfloat{
    \includegraphics[width=0.9\linewidth]{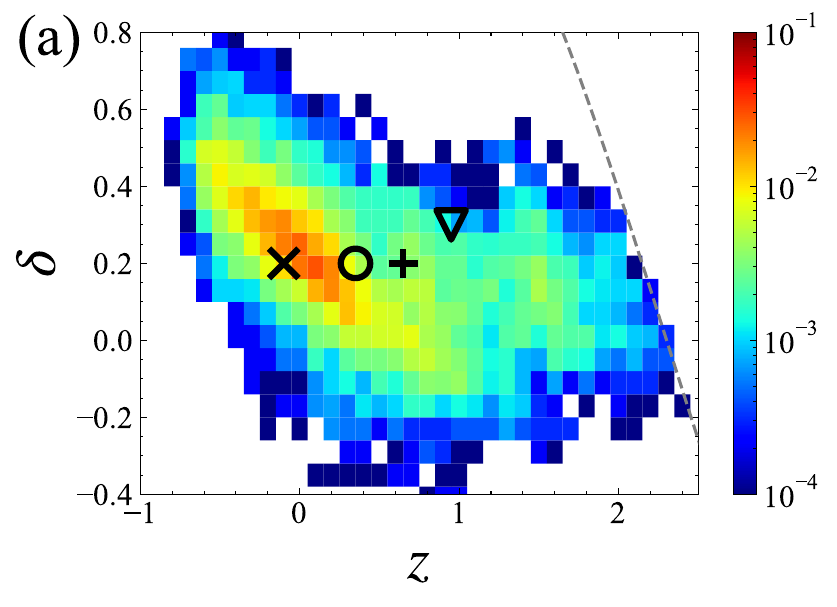}\label{fig:neutron_map_zdelta}
  }
  \\
  \subfloat{
    \includegraphics[width=0.9\linewidth]{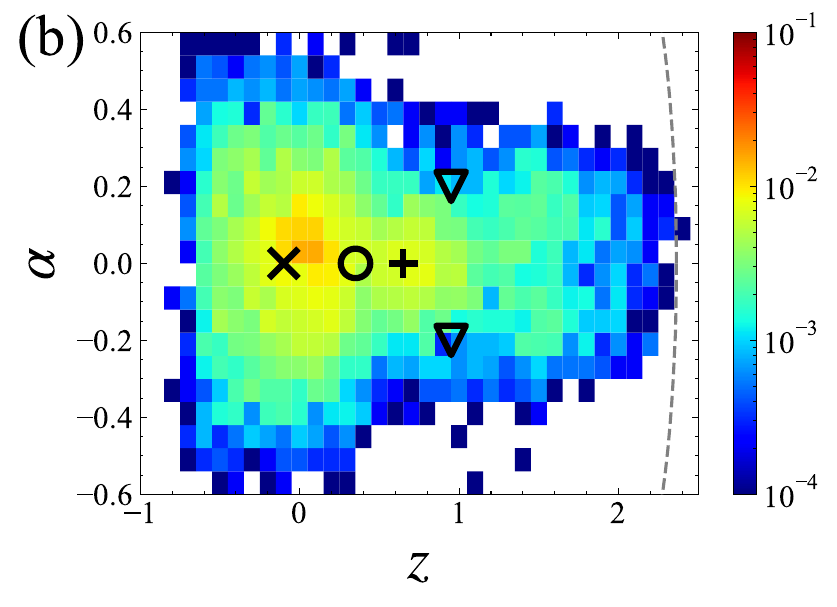}\label{fig:neutron_map_zalpha}
  }
  \caption{
    Probability of neutron emission plotted on the (a) $z-\delta$ and (b) $z-\alpha$ planes for compound nucleus $^{238}\mathrm{U}$ at $E^* = \SI{45}{\MeV}$.
    The ground state and second minimum are represented by $\times$ and $+$, respectively.
    The first and second saddle points are indicated by $\bigcirc$ and $\triangledown$.
    The scission line is shown with the gray dashed line.
  }\label{fig:neutron_map}
\end{figure}

The difference of the isotope dependence for FFMD between the present calculation and the GEF-based calculation \cite{Tanaka2019-yy} at high excitation energy can be interpreted as follows.
First of all, as discussed in \Cref{fig:neutron_map}, most of the neutrons are emitted around the ground-state shape, so that the effect of shell structure is already predominantly restored before overcoming the saddle point.
Thus, difference comes mainly from the total number of emitted neutrons.
For example, in \Cref{fig:average_neutron}, if one examines neutron multiplicity for neptunium isotope at an excitation energy of \SI{55}{\MeV},
it is shown that GEF calculation predicts a change of neutron multiplicity from -24.9 \% to +21.8 \% relative to $\nuclide{239}{Np}$, whereas in the present case the change is limited from -14.0 \% to +10.4 \%.
Thus the present model shows a weaker dependence of the effects of multichance fission with respect to the fissioning-nucleus mass.

\begin{figure}[htbp]
  \centering
  \includegraphics[width=0.9\linewidth]{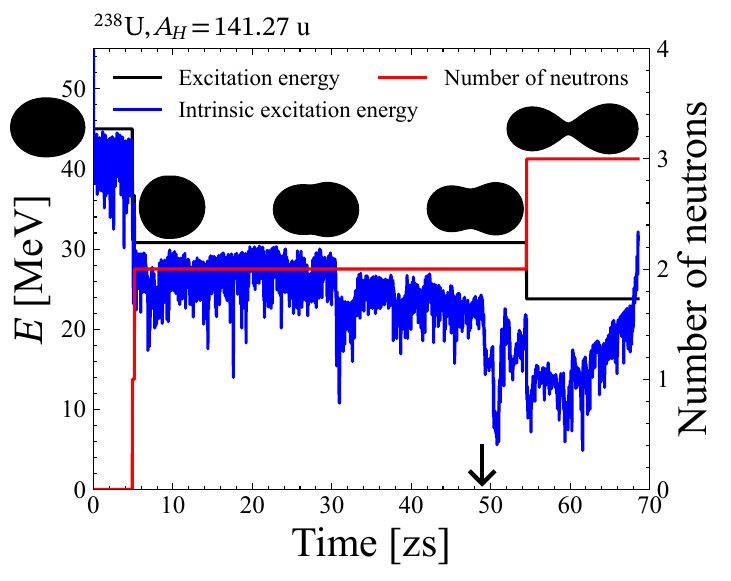}
  \caption{
    Sample trajectory of the time evolution of the excitation energy ($E^*$) (black), the intrinsic excitation energy $E_\mathrm{int}$ (blue) and the number of emitted neutrons (red) for $^{238}\mathrm{U}$ at $E^* = \SI{45}{\MeV}$.
    The black shapes represent the nuclear shape at each point.
    The arrow indicates the time at which the nucleus passes through the second saddle point.
  }\label{fig:ex_time}
\end{figure}

Figure \ref{fig:ex_time} shows an example of the time evolution of the excitation energy, the intrinsic excitation energy ($E_{\mathrm{int}}$), and of the number of emitted neutrons from the compound nucleus $\nuclide{238}{U}$ from an initial excitation energy $E^* = \SI{45}{\MeV}$ to the scission point.
The difference between the excitation energy (black line) and the intrinsic excitation energy (blue) corresponds the collective vibrational energy of the nucleus.
At first, the intrinsic excitation energy is around \SI{40}{MeV}.
At around \SI{5}{zs}, neutron emission occurs and the $E^*$ and $E_\mathrm{int}$ decrease by a magnitude of \SI{6}{\MeV} and \SI{7}{\MeV}, respectively.
A second neutron is then emitted almost immediately.
At \SI{55}{zs}, the third neutron is emitted after the nucleus has overcome the 2nd saddle point and is in the vicinity of the scission point.

\begin{figure}[htbp]
    \centering
    \subfloat{
        \includegraphics[width=0.9\linewidth]{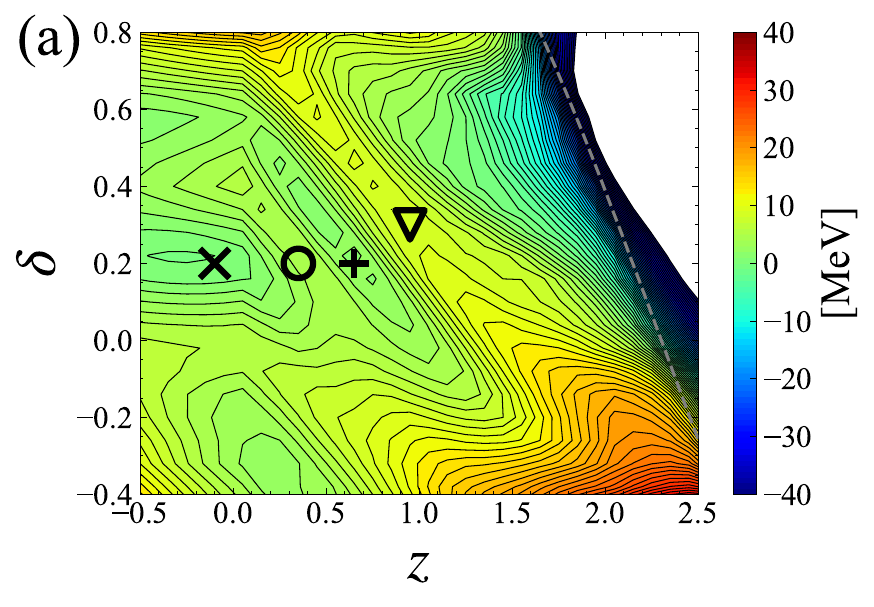}\label{fig:zdelta_pes}
    }
    \\
    \subfloat{
        \includegraphics[width=0.9\linewidth]{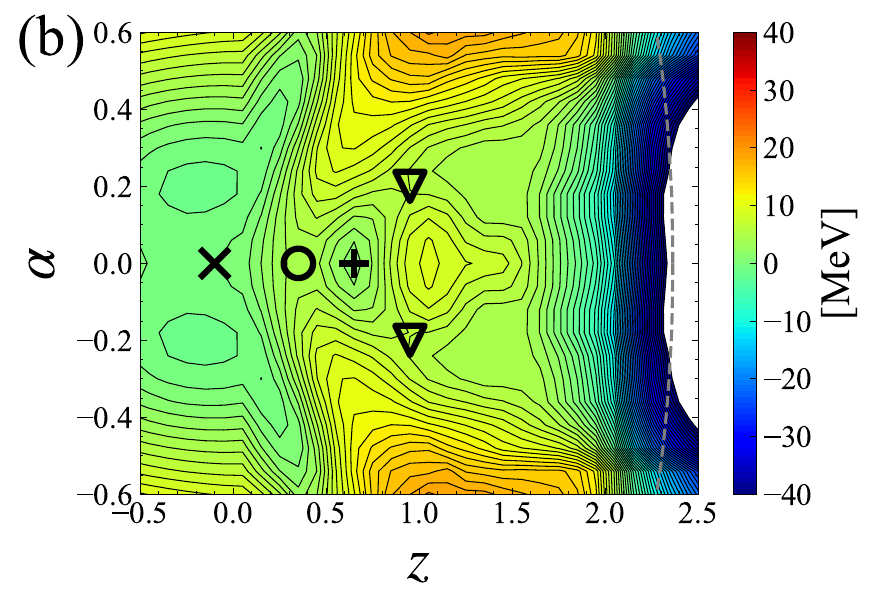}\label{fig:zalpha_pes}
    }
    \caption{
        Potential energy surface on the (a) $z-\delta$ ($\alpha = 0.0$) and (b) $z-\alpha$ ($\delta = 0.2$) planes for $\nuclide{238}{U}$.
        The ground state and second minimum are represented by $\times$ and $+$, respectively.
        The first and second saddle points are indicated by $\bigcirc$ and $\triangledown$.
        The scission line is shown with the gray dashed line.
    }\label{fig:pes}
\end{figure}

Figure \ref{fig:neutron_map} shows the neutron-emission probability distributions plotted on the (a) $z-\delta$ plane and (b) $z-\alpha$ plane, from an initial excitation energy of $E^* = \SI{45}{\MeV}$, for compound nucleus $\nuclide{238}{U}$.
The corresponding potential energy surface is also shown in \Cref{fig:pes}, in which the ground state and second minimum, as well as the first and second barrier positions, are indicated.
As seen in \Cref{fig:neutron_map,fig:pes}, most of the neutrons are emitted near the ground-state shape.
The new approach gives nearly the same results as the general statistical-model calculation, that neutron emission happens at the ground-state shape.
Furthermore, our calculation does not conflict with the observed experimental signature of enhanced neutron-induced fission cross sections emerging at the excitation energy threshold of each fission chance\ \cite{Lestone2011-gn, Tovesson2014-hn}.
An important message from the present calculation is that the second minimum at $z=0.7$ represents another key neutron emission point.
This is because the system is sitting around this shape for a relatively long time.
We also stress that neutrons are emitted at any shape of the nucleus between the saddle point to near the scission point, in the region where $z > 1$.

\section{Summary}
Fission-fragment mass distributions were calculated in the Langevin-equation model for 21 U, Np, and Pu nuclei, and their excitation-energy dependence was obtained from an initial excitation energy $E^*= \SI{15}{\MeV}$ to $\SI{55}{\MeV}$.
Neutron emission is handled through the evolution of nuclear shape from the ground state to the scission point, in contrast to the usual statistical model.
While the results of the calculations with and without neutron emission do not change for low-excitation-energy fission, the difference becomes significant at high excitation energy, where the double-peak structure evident in experimental data is reproduced due to the revival of shell correction energy due to neutron emission.
The present calculation, which addresses neutron emission throughout the fission process, demonstrates that measured FFMDs can be adequately predicted for high-energy fissions,
and thus can be considered a viable alternative to the traditional approach for discussing multichance fission.
Concerning the origin point of neutrons, they are mostly emitted from the ground-state shape.
However, there is an enhancement in neutrons emitted around the second minimum, and between the saddle and scission points.
Thus, the present calculations can be applied to discuss fission time scales by comparison to experimental data of pre-scission neutrons.

\section*{Acknowledgments}
We thank J. Smallcombe of JAEA for the fruitful discussion.
The Langevin calculations were performed using the cluster computer system (Kindai-VOSTOK)
which is supported by the Japan Society for the Promotion of Science (JSPS) KAKENHI Grant Number 20K04003 and 24K07060.
One of the authors (S.T.) thanks the support from the JAEA special research student program.

\bibliography{mcf}

\end{document}